# Research on the pixel-based and object-oriented methods of urban feature extraction with GF-2 remote-sensing images

**Dong-dong Zhang[1], Lei Zhang[1,2*], Vladimir Zaborovsky[3], Feng Xie[4], Yan-wen Wu[1], and Ting-ting Lu[1]**

1 Shanghai Key Laboratory of Multidimensional Information Processing, East China Normal University, Shanghai, China
2 MOE International Joint Lab of Trustworthy Software, East China Normal University, Shanghai, China
3 Department of Computer Systems and Software Engineering, Peter the Great St. Petersburg Polytechnic University, St. Petersburg, Russia
4 Institute of Technical Physics, Chinese Academy of Sciences, Shanghai, China

**\*** Correspondence: lzhang@ce.ecnu.edu.cn

**Abstract:** During the rapid urbanization construction of China, acquisition of urban geographic information and timely data updating are important and fundamental tasks for the refined management of cities. With the development of domestic remote sensing technology, the application of Gaofen-2 (GF-2) high-resolution remote sensing images can greatly improve the accuracy of information extraction. This paper introduces an approach using object-oriented classification methods for urban feature extraction based on GF-2 satellite data. A combination of spectral, spatial attributes and membership functions was employed for mapping the urban features of Qinhuai District, Nanjing. The data preprocessing is carried out by ENVI software, and the subsequent data is exported into the eCognition software for object-oriented classification and extraction of urban feature information. Finally, the obtained raster image classification results are vectorized using the ARCGIS software, and the vector graphics are stored in the library, which can be used for further analysis and modeling. Accuracy assessment was performed using ground truth data acquired by visual interpretation and from other reliable secondary data sources. Compared with the result of pixel-based supervised (neural net) classification, the developed object-oriented method can significantly improve extraction accuracy, and after manual interpretation, an overall accuracy of 95.44% can be achieved, with a Kappa coefficient of 0.9405, which objectively confirmed the superiority of the object-oriented method and the feasibility of the utilization of GF-2 satellite data.

**Keywords:** Image classification; GF-2 satellite data; Object-oriented; Urban mapping; Accuracy assessment

# 1. Introduction

Nowadays, sustainable development has been a global goal and has been promoted by the joint international Land Use/Cover Change project which is carried out by the International Geosphere – Biosphere Programme (IGBP) and the International Human Dimensions Programme on Global Environmental Change (IHDP) [1]. Detecting and monitoring land-use change have attracted many research interests [2, 3]. Substantial work has been done with regard to land-use classification modeling over the past few decades [4-6]. Various models have been developed to obtain accurate and detailed information of urban ecological land for both urban landscape ecosystem construction and urban land management analysis [7, 8]. Mastering the spatial distribution of urban landform information and its changing characteristics will help people's decision-making and promote the development planning of the city [9].

Over the past few decades, experts and scholars all over the world have been focusing on remote sensing data classification methods, from supervised and unsupervised classifications based on traditional statistical analysis to widely used support vector machine [10], neural network [11] and expert decision system [12]. Among them, the pixel-based statistical classification technique has been dominant and relatively mature, and has achieved good application results in some fields [13-15]. However, the conventional pixel-based classification methods only utilize spectral information and consequently have limited success in classifying high-resolution multispectral urban images with similar spectra between different categories [16]. Spatial information such as geometric and spatial features must be exploited to produce more accurate classification maps [17]. In recent years, object-oriented classification methods have drawn tremendous research interests [18-20]. It is proven to have the potential to overcome the weaknesses associated with per-pixel analysis, for instance, negligence of geometric and contextual information [21, 22]. The main idea is to first segment the image into objects with certain meanings, and then to classify it by using the spectral, shape and neighborhood features of the objects. This method takes more discriminative features into consideration and adapts to the human visual interpretation habits, which provides a new way of thinking for information extraction.

Both the reference demand for urban construction decision-making and the supervision of urban land objects depend on the remote sensing data acquisition of urban surface coverage information. With the advent of high-resolution satellites, remote sensing image systems developed by different countries have provided more and more spatial texture information for urban planning and construction, such as Rapid-Eye [23] (with a resolution of 5.0 meters) in Germany, SPOT5 [3] (2.5 meters) in France, QuickBird [24] (0.61 meters) and World-View series [25] (0.5 meters) in the United States. GF-2 satellite is a recently developed civil optical remote sensing satellite with sub-meter spatial resolution, which has the characteristics of high spatial resolution, high positioning accuracy and fast attitude maneuverability [26]. It is significant to study how to make full use of the advantages of GF-2 remote sensing image data to extract urban surface information, and find out whether it is more value-oriented from domestic data self-sufficiency, or meets the needs of urbanization development with good timeliness and excellent quality [27]. At present, the urbanization development of China is rapid. Hence, how to ensure the harmonious development of environment and urbanization is an urgent problem to be solved. In this work, it is of great significance to obtain accurate and detailed information of urban ecological land-use for both urban landscape ecosystem construction and urban land management analysis [28]. In this paper, the eCognition software [29] is used to explore the acquisition method of optimal segmentation scale. Through the spectral and spatial analysis of image objects, the system rules of object extraction are studied, and a set of typical object-oriented urban object extraction methods based on GF-2 are established.

## 2. Studied Area and Methodology

### 2.1 *Studied Area*

In order to explore the universal applicability of classification methods, the selected area to be studied is the whole Qinhuai District, the central area of Nanjing City, which is named after Qinhuai River running through the whole territory, as shown in Figure 1. Qinhuai District is a hilly area, showing a terrain which is high in the southeast and low in the northwest. The landform is dominated by plains. There are numerous surface water systems, mainly including the Inner Qinhuai River and its branches, and scattered natural and artificial rivers, forming the wavy topographic landscape. The ground elevation is between 6 and 12 m. Qinhuai District has a subtropical humid climate with four distinct seasons and abundant rainfall.

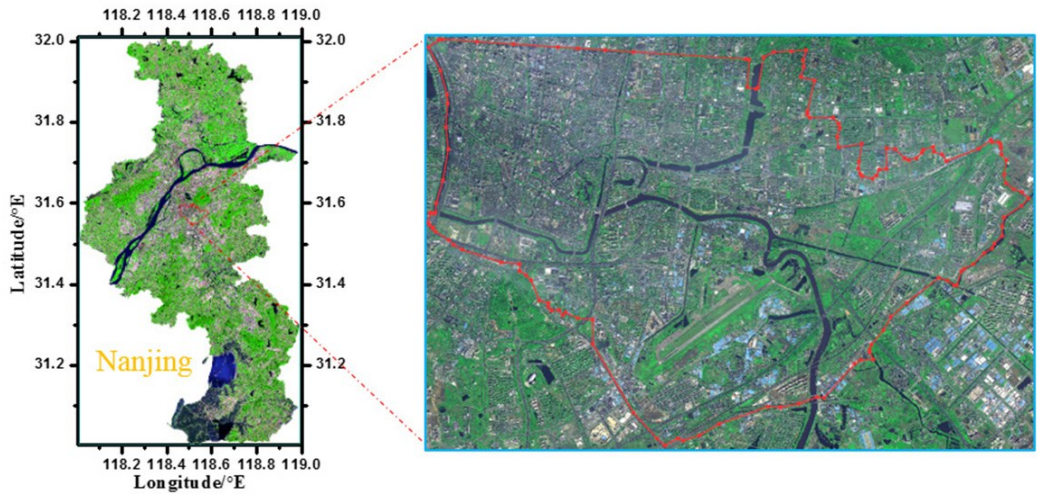

Figure 1. Location of the studied area - Qinhuai District

## 2.2 *Data Acquisition and Preprocessing*

A cloud-free GF-2 satellite image was obtained from CRESDA after geometric correction with default WGS-84 coordinate system. The satellite data contains panchromatic (PAN) data file, multispectral (MS) data file and RPC file which has detailed RPC model parameters to be used for image orthorectification. The initial data was taken on April 21, 2015 covering the central area of Nanjing. Nanjing is located in the central region of the lower reaches of the Yangtze River and southwest of Jiangsu Province, the longitude ranges from 118°22′ E to 119°14′ E and latitude ranges from 31°14′ N to 32°37′ N. The geographic location of obtained data is shown in Fig. 1.

The dashed box in Figure 2 shows the preprocessing flow of the raw GF-2 remote sensing image data which includes radiometric correction, orthorectification, atmospheric correction, image fusion and cropping. Radiometric correction corrects all the radiation-related errors in order to eliminate interference and obtain the real reflectivity data. Orthorectification corrects the image point errors caused by terrain fluctuations, and atmospheric correction eliminates the mist in the air and makes the image clearer and brighter. Meanwhile, the resolution of the MS image and PAN image are resampled to 4 meters and 1 meter during the process of orthophoto correction, respectively. The PAN and MS bands are fused using the Nearest Neighbor Diffusion (NNDiffuse) pan sharpening algorithm, which can achieve good spectral fidelity while maintaining high spatial detail and then cropped with a drawn vector boarder file to get the studied area. The whole preprocess of GF-2 image is carried out in ENVI. In addition, the drawing of above-mentioned vector boundary file and vector sample point files and the calculation of the classification results are performed in ArcGIS.

## 2.3 *Methodology*

According to the characteristics of GF-2 remote sensing image and research objectives, combined with actual survey data and 2015 land-use data, the urban land types are divided into five categories: Vegetation, Water, Road, Bare Land and Building. The classification was carried out initially by using the neural net classification algorithm [30]. The accuracy assessment was carried out using ground truth data. Then, multi-scale segmentation and establishment of various feature rules for object-oriented extraction of urban feature information were performed in eCognition software in order to establish a classification method of rule set for the study area of GF-2 image. Finally, the object-oriented classification results were compared with the traditional supervised classification results, and the accuracy analysis and mapping were performed. The flow chart of the classification technique is shown in Fig. 2.

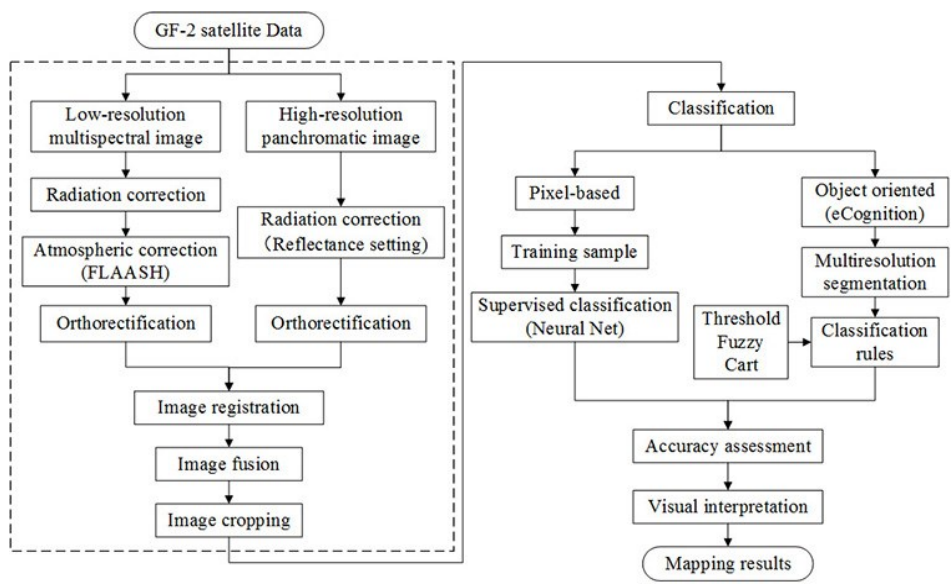

Figure 2. Classification technique flow chart

## 3. Object oriented classification

Object-oriented classification can outperform the traditional pixel-based classification method. Instead of using pixels as the minimum unit, it divides the image into objects and uses the spectral and spatial features between the objects to classify them [31]. The core process consists of two steps, namely segmentation and classification.

### 3.1 *Multiresolution segmentation*

Image segmentation is the basis of object-oriented information extraction, and the efficiency of segmentation directly determines the quality of information extraction results [32]. Here the multiresolution segmentation technique is used for image segmentation, which is realized based on the region merging technique under the premise that the average heterogeneity is the smallest and the homogeneity between the internal pixels of the object is the largest. In multi-scale segmentation, the selection of scale is particularly important for segmentation: too large scale makes the image under-segmented, since the same object contains a variety of terrain information, and too small scale makes the segmented object too fragmented, which is difficult to manipulate [33]. Due to the large area of the studied area and the complexity of the terrain features, it is not suitable for establishing a segmentation hierarchy. For road segmentation, a larger segmentation can be chosen for suburban roads, but urban roads are affected by vegetation and buildings on both sides, which will lead to the segmented objects containing other feature information, thereby affecting the classification accuracy. The bands participating in image segmentation are Blue, Green, Red, and Near-infrared (NIR), and the weight of each is 1. The optimal values of scale parameter, shape/color and compactness/smoothness are selected accordingly to optimize the image segmentation results. After many tests, the following multiresolution segmentation parameters were used for extracting different urban classes: scale parameter = 100, shape = 0.2, and compactness = 0.6, which leads to optimal results consistent with that of ESP (estimation of scale parameter) tool [34]. Partial segmentation results are shown in Figure 3.

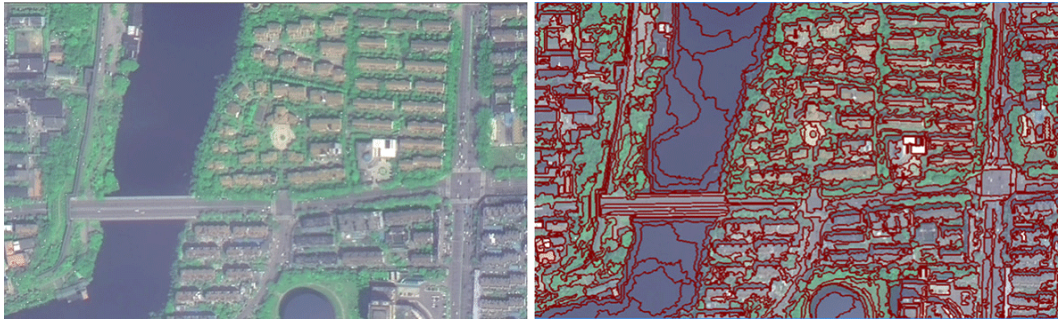

Figure 3. Partial segmentation results

## 3.2 *Establishment of classification rules*

Multiresolution segmentation was followed by classification of image objects. The object-oriented classification technique adopts the fuzzy classification algorithm supported by the decision expert system, making full use of the spectrum, shape, texture, context, spatial relationship and other characteristics of the image, which breaks the limitations of traditional image classification based on spectral information [35].

In this section, two object-oriented classification methods are performed. One is the combination of threshold classification and fuzzy classification. The RGB mode was used to combine the bands for display while using the near infrared band for enhancement. Vegetation, Water, Road, Bare Land and Building are extracted and classified successively. By using the spectral, geometry, position and texture feature information of each object in the image, selecting the appropriate classification features and forming the feature combination of the extracted feature categories, a classification rule set of various features is established, as shown in Table 1. The fuzzy functions being used to classify different object features of GF-2 image are shown in Fig. 4. A decision sequence for the extraction of ground information is necessary in order to ensure accuracy and improve information extraction efficiency. First, the types of features that are easier to extract are extracted, and then the types of features that are relatively difficult to distinguish are extracted. Vegetation information is preferentially extracted in the extraction process for its obvious difference in spectral characteristics compared with other land types. A user-defined feature, Normalized Difference Vegetation Index (NDVI) is used to distinguish vegetation from other land types. NDVI >0.22 is determined as vegetation and the value of NDVI between 0.16 and 0.22 may be vegetation and other features, so a fuzzy function is used for the fuzzy interval, as shown in Fig. 4. The NDVI is defined as:

$$\text{NDVI} = \frac{Mean\ NIR - Mean\ Red}{Mean\ NIR + Mean\ Red} \qquad (1)$$

where $Mean\ NIR$ respects the values of near infrared band, $Mean\ Red$ respects the values of Red band in the GF-2 Image. However, by using this classification index, blue roof buildings will be misclassified. This kind of misclassification is prevented by using the spectral feature: Mean Blue >1250 and those buildings are classified as *buiding1*. The urban land cover information is divided into vegetation and non-vegetation and subsequent feature categories are extracted from non-vegetation.

Compared with other features, Water has obvious difference and particularity in spectral feature and texture feature, which can be clearly distinguished. However, the characteristics of the GF-2 data are similar to that of the shadow information and the water body information, resulting in a large amount of information mixed with shadow information in the extracted water body information. For the studied area, most of the shaded areas belong to the category of building. In order to minimize the interference of shadow areas, Normalized Difference Water Index (NDWI) [36] and NIR are used for water classification, and many tests have proved that combining NDWI and NIR can achieve better classification results than solely using either of them. The NDWI is defined as:

$$\text{NDWI} = \frac{Mean\ Green - Mean\ NIR}{Mean\ Green + Mean\ NIR} \qquad (2)$$

where $Mean\ Green$ represents the values of Green band, and $Mean\ NIR$ respects the values of NIR band. Before the road classification, Area < 60 pixel is set to exclude the interference of vehicles and merge these objects into unclassified parts. Road usually has a large aspect ratio (length/width) and is strip-shaped, which can be constrained by shape index. However, due to the building style of Qinhuai district, some roofs of Building have similar geometry features, and therefore spectral features are used to distinguish these different feature categories.

After Road information is extracted, the remaining unclassified areas include the Bare Land and Building information. Bare Land includes unused bare ground, area of sparse vegetation, exposed areas on both sides of roads and rivers, construction sites to be built, and exposed land that has not been disposed of in time after construction, etc. Due to its scattered distribution and large spectral range, Red and NIR spectral features, Brightness and Asymmetry are used to extract bare land information. Building, that is, construction land, includes the remaining part after the object classes above are extracted and also the separated building area.

Table 1. Classification rule set

| Typical urban features | Classification rule |
|---|---|
| Vegetation | Membership function: NDVI, fuzzy interval: (0.16, 0.22); 0.22 ≤ NDVI; Remove: Vegetation with Mean Blue > 1250 (building1); |
| Water | -0.085 ≤ NDWI and Mean NIR < 1100; |
| Road | Area < 60 pixel; Length/Width > 4.8 and shape index > 2; Remove: Road with Mean Blue > 1450 (building2); Road with Mean NIR > 1930 and Mean Blue < 1200 (building3); |
| Bare Land | Membership function: Asymmetry, Fuzzy interval: (0, 0.8); Brightness, fuzzy interval: (1050, 1250); Mean Red, fuzzy interval: (1180, 1500); NIR, fuzzy interval: (1380, 1800); |
| Building | The remaining part after the object classes above are extracted; building1, building2, building3; |

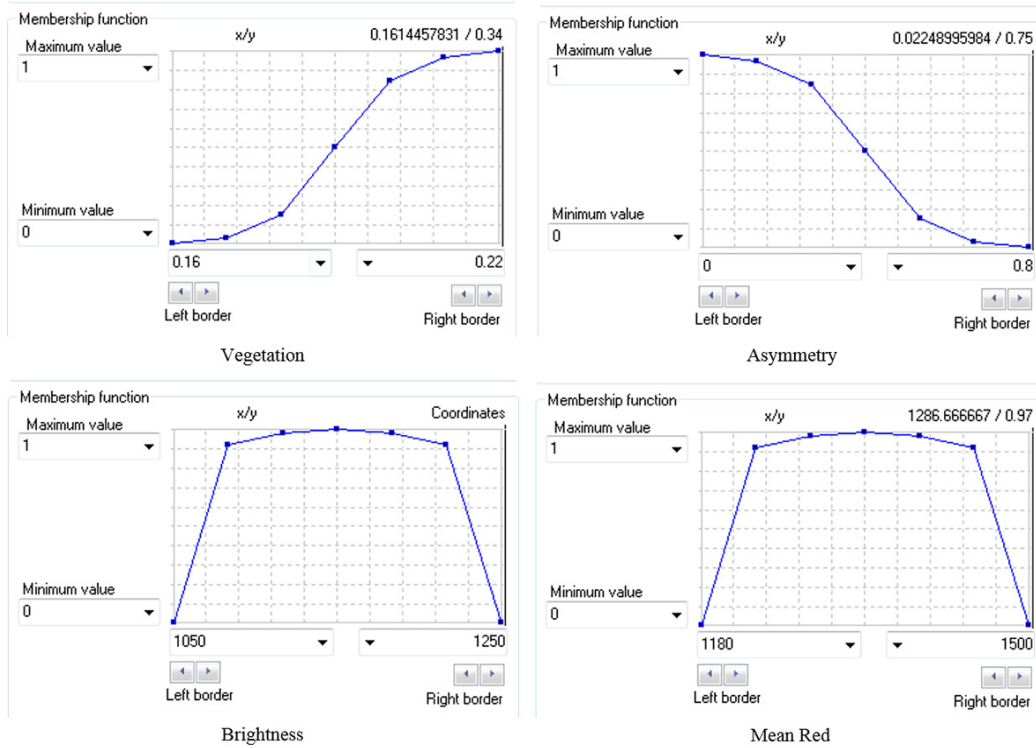

Figure 4. Membership function for selected classes

Another object-oriented classification is Cart decision tree classification method which can be used for big data image classification. A vector sample point file is drawn by visual interpretation in ArcMap software with category names in the vector data. Then the vector file is exported into eCognition and converted into samples according to the vector classification object. And finally, the image is classified by Cart algorithm without manually finding the best feature for extracting the category.

## 4. Results and Discussion

### 4.1 *Classification Results*

Five typical urban features are finally extracted by combining threshold and fuzzy classification methods according to the established classification rule set in Table 1. The classification result is shown in Fig. 5, and the classification result with Cart decision tree classification method is shown in Fig. 6. The classification results shown in Fig. 5 and Fig. 6 are the original extracted data that have not been manually interpreted.

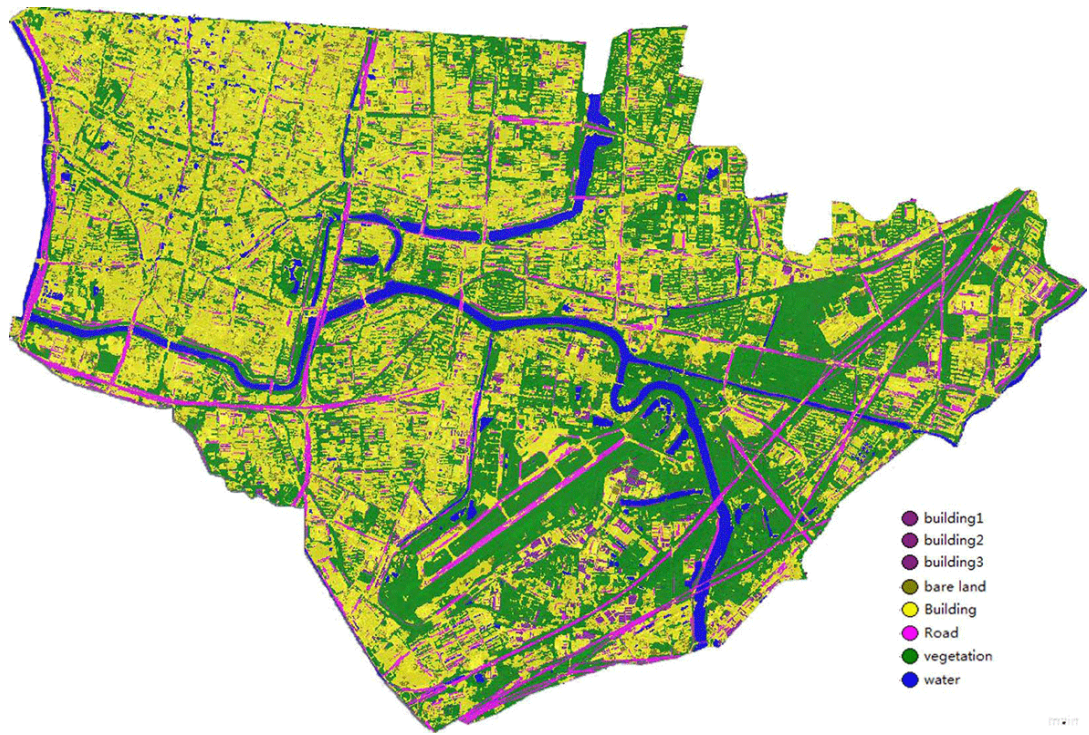

Figure 5. Threshold and fuzzy classification result

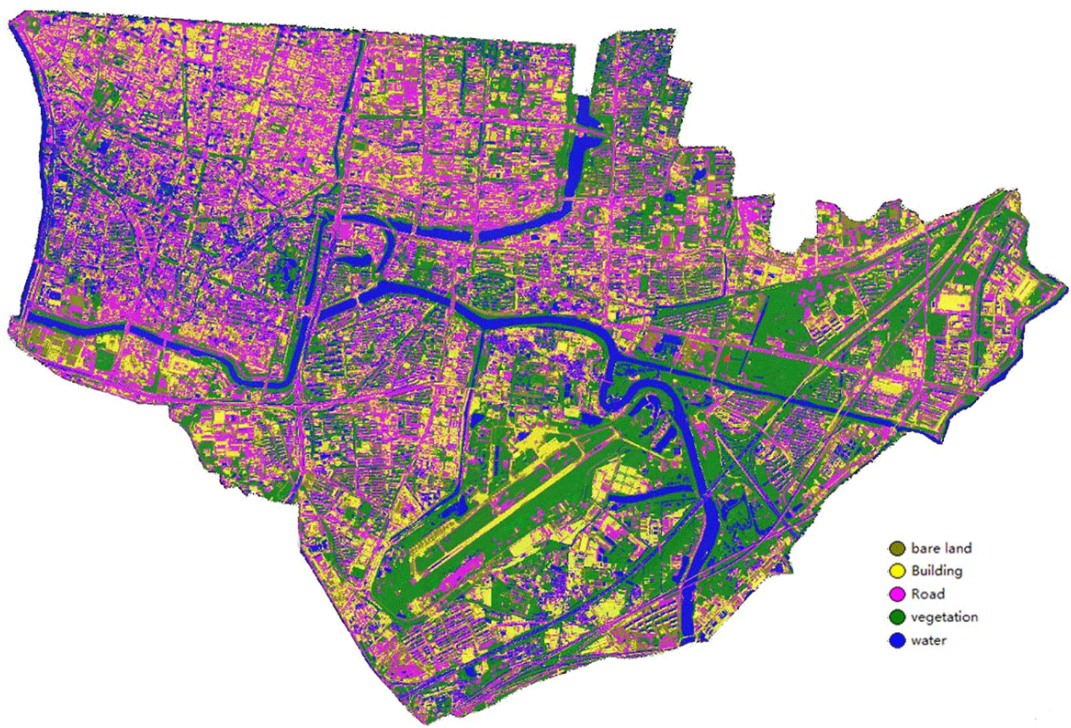

Figure 6. Cart decision tree classification result

It is unrealistic to use only one of the established classification rules to completely extract all types of features. The establishment of classification rules only meets the needs of most terrestrial information extraction and classification. The classification rules are not applicable for some specific features, such as the shadow regions which might include all the feature categories and water bodies where phytoplankton exist. It should also be noted that more features involved in classification do not mean higher accuracy of classification, because more classification features may also cause the phenomenon of feature redundancy, which increases the amount of calculation, reduces the classification efficiency, and even reduces the classification accuracy. Hence, manual interpretation is indispensable. For comparison, visual interpretation is carried out based on Google Earth data and the result is shown in Fig. 7.

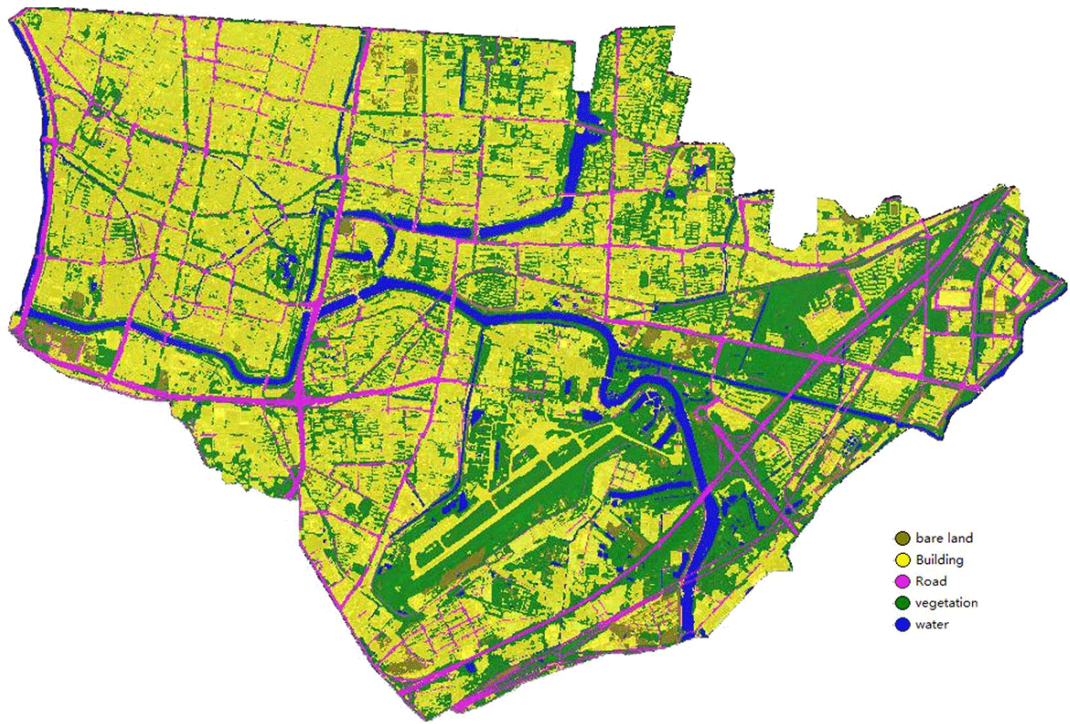

Figure 7. Visual interpretation results

The above-mentioned classification results were exported into ArcMap to count category area and the statistical results are shown in Table 2. The proportions of the final classification results are relatively close to that of the 2016 Statistical Yearbook in Nanjing. The differences between the former classification result and the visually interpreted result are mainly in the categories of Building and Vegetation. The possible reasons of that might be: 1) Qinhuai district is rich in vegetation and many buildings are surrounded by vegetation, and hence, their spectral information is disturbed, making the spectral feature biased towards vegetation information, which results in the increase of vegetation area ratio and the decrease of building area ratio; 2) parts of some special sites, such as airport, sports ground and building site, are misclassified as Road while they should be considered as Building; 3) plastic venues including basketball courts and tennis courts, or green roof buildings, are mistaken for vegetation; 4) some exposed ground surfaces at building sites are covered to prevent dust pollution, which are considered as Building. The latter classification result by using Cart decision tree classification method has a large difference with the visually interpreted result, mainly reflected in Building, Road and Water. Road has high similarity in spectral and shape features with gray roof buildings, which makes it hard to differentiate Road from Building. Water and the shadow information have high similarity and are easily confused due to the characteristic of GF-2 data, resulting in the misclassification of Water and Building, especially in compact building districts in the north-western Qinhuai district. The misclassification of dark roof construction for water is also because of the spectral similarity.

Compared with the classification results of the spatial distribution and the statistical data, the former object-oriented classification result is closer to the final visual interpretation result, which indicates that the efficiency of category extraction by combining threshold and fuzzy classification is better than that of Cart decision tree classification method for the studied area of GF-2 image.

Table 2. Statistics results of classification

| Classification result | Class name | Perimeter (km) | Area (ha) | Area ratio |
|---|---|---|---|---|
| Threshold and fuzzy classification result in Fig. 5 | Bare Land | 919.06 | 195.95 | 3.98 |
| | Building | 11419.29 | 2294.53 | 46.59 |
| | Road | 2147.02 | 391.74 | 7.95 |
| | Vegetation | 8892.68 | 1817.19 | 36.9 |
| | Water | 463.10 | 225.43 | 4.58 |
| Cart decision tree classification result in Fig. 6 | Bare Land | 2170.77 | 429.12 | 8.06 |
| | Building | 5590.48 | 1033.65 | 24.64 |
| | Road | 6991.07 | 1421.55 | 25.69 |
| | Vegetation | 5993.53 | 1316.00 | 26.33 |
| | Water | 3030.88 | 724.52 | 14.71 |
| Visual interpretation results in Fig. 7 | Bare Land | 475.37 | 124.08 | 2.52 |
|  | Building | 12760.66 | 2622.30 | 53.25 |
|  | Road | 2162.97 | 422.09 | 8.57 |
|  | Vegetation | 6895.63 | 1505.17 | 30.56 |
|  | Water | 631.40 | 251.20 | 5.1 |
|  |  |  | 4924.84 |  |

## 4.2 *Accuracy assessment*

Sample points are selected by referring to the pan-sharpening GF-2 image in ArcGIS to validate the accuracy of pixel-based and object-based classification methods, and then are converted to a raster format, which is considered as the ground truth samples. Neural net classification is a pixel-based supervised classification method which has been more researched and found to have better classification results in recent years and hence, is used for comparison. The accuracy assessment results include three parts, namely the confusion matrix, the accuracy analysis results of a single category, and the accuracy analysis results of the overall category. The confusion matrix is the accuracy assessment matrix generated by using ground truth samples, which is a two-dimensional matrix formed by the actual categories and the classification categories. Accuracy evaluation indexes such as user accuracy, producer accuracy, overall accuracy, and kappa coefficient can be obtained through the confusion matrix. The results of accuracy assessment are shown in Table 3.

Table 3. Pixel-based and object-based classification results of accuracy assessment

| Category Accuracy (%) | Classification method | Vegetation | Road | Building | Water | Bare Land |
|---|---|---|---|---|---|---|
| Producer | (Object-based) Threshold and fuzzy classification | 100 | 62.84 | 81.5 | 79.35 | 23.86 |
| User |  | 93.92 | 78.23 | 66.32 | 80.4 | 63.63 |
| Overall |  | 78.56 ||||
| Kappa coefficient |  | 0.7132 ||||
| Producer |  | 98.31 | 89.74 | 98.7 | 96.42 | 97.3 |

| User | visual interpretation based on object - oriented classification | 96.9 | 99.43 | 87.7 | 99.75 | 98.31 |
| :---: | :---: | :---: | :---: | :---: | :---: | :---: |
| Overall | | 95.44 | | | | |
| Kappa coefficient | | 0.9405 | | | | |
| Producer | (Pixel-based) Neural net classification | 87.22 | 26.28 | 63.46 | 96.56 | 56.15 |
| User | | 84.57 | 19.24 | 82.38 | 43.73 | 13.92 |
| Overall | | 64.37 | | | | |
| Kappa coefficient | | 0.4889 | | | | |

The object-based threshold and fuzzy classification method shows better assessment result than pixel-based neural net classification according to Table 3. Overall classification accuracy can reach 95.44% and Kappa coefficient can reach 0.9405 after manual interpretation. For threshold and fuzzy classification, Vegetation shows the best extraction efficiency, followed by Water, Building and Road. Bare Land shows a larger omission which needs further improvement on feature extraction, but the overall classification accuracy is less affected due to the small proportion of this category. However, the extraction efficiency of different feature categories is much better than that with neural net classification method. Besides, the errors caused by objective factors are inevitable and are summarized as follows: 1) there is a deviation in the boundary vector file drawing which results in the deviation of feature category information but is small; 2) GF-2 data have a small number of bands and insufficient spectral information. The spectral information, i.e. "same spectrum with different objects", affects the classification result, especially for the pixel-level classification method. This is also the reason why pixel-based classification method shows poor accuracy assessment for ground object information extraction.

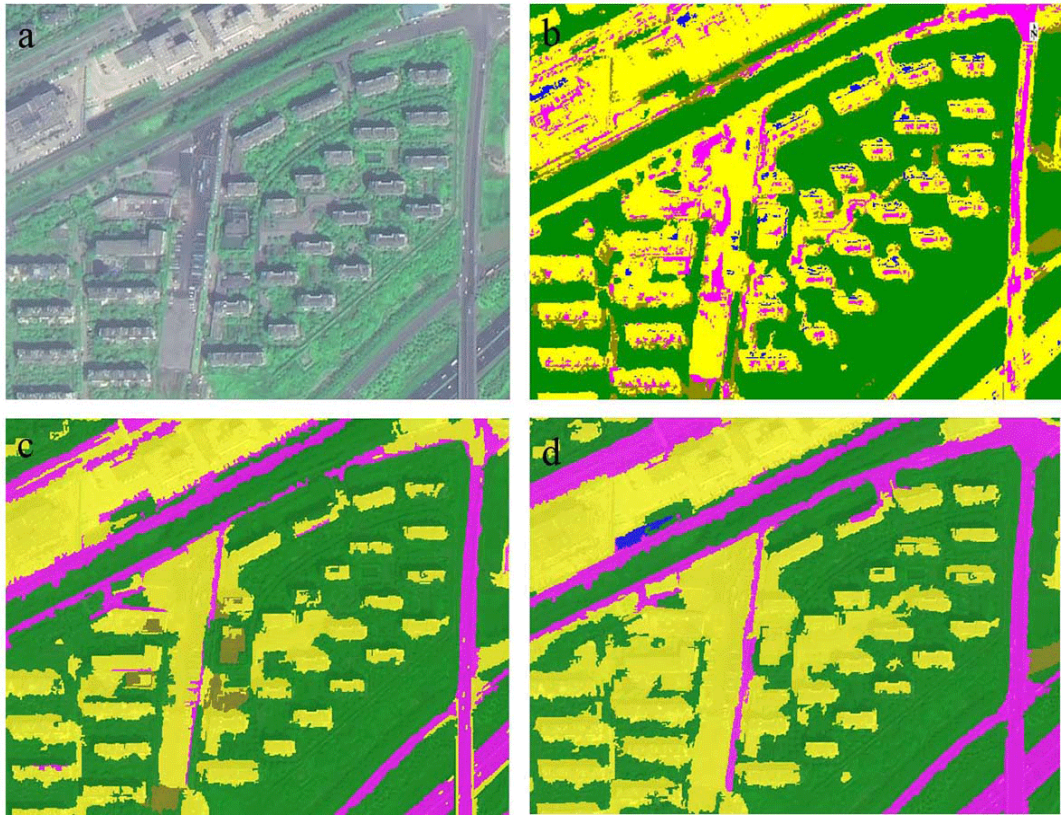

Figure 8. Partial enlargement of classification results: a. Raw image; b. Neural network classification; c. Threshold and fuzzy classification; d. Manual interpretation

For the studied area, the object-oriented classification and the pixel supervised classification result maps are partially enlarged, as shown in Fig. 8. For neural net classification, there are a large number of mixed pixels affecting the classification, resulting in serious noise in the classification results, salt and pepper phenomenon, poor visual effect, and unreasonable classification results. Therefore, the classification result vector is difficult to use, as shown in Figure 8(b). The superiority of object-oriented classification method for urban feature information extraction is objectively proved.

## 5 Conclusions

This work employs the object-oriented classification method to study the extraction of urban feature information based on the GF-2 high-resolution remote sensing images. Through the exploration of different object-oriented classification methods, a technical route suitable for data extraction was designed combining the data characteristics of GF-2 remote sensing images. The remote sensing image of GF-2 in Qinhuai District of Nanjing was used as the researched area to perform the experiment of urban feature information extraction, which achieved the goal of high-efficiency and high-precision information extraction.

According to the information of main urban objects (Vegetation, Water, Building, Road, and Bare Land), the decision-making scheme of information extraction based on the difficulty extraction sequence is proposed, which largely avoids misclassification caused by the similar spectral features and similar shape features between different land types and effectively improve the classification accuracy. The overall accuracy is 95.44% and the kappa coefficient is 0.9405. The superiority of the object-oriented method and the feasibility of the research method are objectively confirmed, which may pave a way for classification and subsequent application research of domestic GF-2 remote sensing images.

**Acknowledgments:** This work is supported by National Key R&D Program of China (No. 2017YFE0100700) and Open Fund of Key Laboratory of Geographic Information Science (Ministry of Education), East China Normal University (Grant No. KLGIS2018A01). I gratefully thank to my research group. I would also like to thank all the members in our international research group who have contributed to the success of our research, either directly or indirectly.

**Author Contributions:** D.Z. conceived and designed the research, processed the data, and wrote the manuscript. Y.W. and T. L. conducted the fieldwork and L.Z. reviewed the manuscript.
**Funding:** This work was supported by the National Key R&D Program of China (2017YFE0100700).
**Conflicts of Interest:** The authors declare no conflict of interest.